\documentclass[aps,pra,twocolumn,showpacs,superscriptaddress,tightenlines,preprintnumbers,sort&compass,amsmath,amssymb,floatfix,nofootinbib]{revtex4}
\usepackage{hyperref}
\usepackage{amsmath}
\usepackage{graphicx}
\usepackage{amssymb}
\usepackage{epsfig}
\usepackage{amsfonts}
\usepackage{color}

\usepackage[utf8x]{inputenc}
\usepackage[T1]{fontenc}

\newcommand{\Wuniv}{\hat W}
\newcommand{\etal}{{\em et al.}}

\newcommand{\<}{\langle} \def\>{\rangle}
 
\newcommand{\tr}{\mathrm{tr}}

\newcommand{\pla}{Phys. Lett. A~}

\def\extra#1{{``#1''}}

\begin{document}

\title{Quantifying entanglement of a two-qubit system via measurable and invariant moments of its
partially transposed density matrix}

\author{Karol Bartkiewicz}
\email{bark@amu.edu.pl} \affiliation{Faculty of Physics, Adam
Mickiewicz University, PL-61-614 Pozna\'n, Poland}
\affiliation{RCPTM, Joint Laboratory of Optics of Palack\'y
University and Institute of Physics of Academy of Sciences of the
Czech Republic, 17. listopadu 12, 772 07 Olomouc, Czech Republic }

\author{Jiří Beran}
\affiliation{RCPTM, Joint Laboratory of Optics of Palack\'y
University and Institute of Physics of Academy of Sciences of the
Czech Republic, 17. listopadu 12, 772 07 Olomouc, Czech Republic }

\author{Karel Lemr}
\affiliation{RCPTM, Joint Laboratory of Optics of Palack\'y
University and Institute of Physics of Academy of Sciences of the
Czech Republic, 17. listopadu 12, 772 07 Olomouc, Czech Republic }

\author{Micha\l{} Norek}
\affiliation{Faculty of Physics, Adam Mickiewicz University,
PL-61-614 Pozna\'n, Poland}

\author{Adam Miranowicz}
\affiliation{Faculty of Physics, Adam Mickiewicz University,
PL-61-614 Pozna\'n, Poland}

\begin{abstract}
We describe a direct method to determine the negativity of an
arbitrary two-qubit state in experiments. The method is derived by
analyzing the relation between the purity, negativity, and a
universal entanglement witness for two-qubit entanglement. We show
how the negativity of a two-qubit state can be calculated from
just three experimentally-accessible moments of the
partially-transposed density matrix of a two-photon state.
Moreover, we show that the negativity can be given as a function
of only six invariants, which are linear combinations of nine
invariants from the complete set of 21 fundamental and independent
two-qubit invariants. We analyze the relation between these
moments and the concurrence for some classes of two-qubit states
(including the $X$~states, as well as pure states affected by the
amplitude-damping and phase-damping channels). We also discuss the
possibility of using the universal entanglement witness as an
entanglement measure for various classes of two-qubit states.
Moreover, we analyze how noise affects the estimation of
entanglement via this witness.
\end{abstract}


% 03.67.Mn Entanglement measures, witnesses, and other characterizations
% 42.50.Dv Quantum state engineering and measurements in quantum optics
% 89.70.+c Information theory and communication theory

\date{\today}
\maketitle

%------------------------------------------------------------------
%%%%%%%%%%%%%%%%%%%%%%%
\section{Introduction}
%%%%%%%%%%%%%%%%%%%%%%%%%
Quantum entanglement~\cite{Schrodinger35,Einstein35}, which is an
intrinsically and fundamentally nonclassical effect, has attracted
an enormous number of works related to quantum information
processing and quantum engineering in the last two decades (for
reviews see, e.g.,
Refs.~\cite{BengtssonBook,Horodecki09,SchleichBook}). Although,
our understanding of quantum entanglement is much deeper now,
there are still many open fundamental problems as listed, e.g., in
Ref.~\cite{WernerList}. Some of these problems address the
question how to experimentally detect and estimate entanglement of
a given state~\cite{Guhne09}.

One could think that the most natural and simplest way to measure
the entanglement of an unknown state of $\rho$ is to apply quantum
state tomography (QST). This approach enables the reconstruction
of $\rho$ by postprocessing experimental data, and, then, the
calculation of arbitrary entanglement measures for $\rho$. Indeed,
various effective QST methods have been
developed~\cite{ParisBook}, including those for the reconstruction
of the polarization states of two photons (for a recent comparison
see Ref.~\cite{Miran14}). Nevertheless, this complete
reconstruction requires to measure also a large number of
parameters, which are irrelevant for the determination of
entanglement. This number scales with the square of the dimension
of a measured state $\rho$. Moreover, QST based on linear
inversion often leads to unphysical reconstructed states. Then,
nonlinear methods (based, e.g., on maximum-likelihood estimation)
have to be applied to overcome this problem.

Thus, usually, entanglement is detected and quantified by
measuring entanglement witnesses~\cite{Horodecki96} (for a review
see Ref.~\cite{Horodecki09}). This approach corresponds to testing
the violations of classical inequalities. The operational
usefulness of entanglement witnesses has been demonstrated in
numerous experimental (see, e.g., earlier experiments reported in
Refs.~\cite{Bourennane04,Haffner05,Leibfried05,Bovino05}) and
theoretical works. The latter include approaches based on:
polynomial moments~\cite{Horodecki02,Horodecki03,Carteret05},
collective entanglement
witnesses~\cite{Horodecki03,Aolita06,Walborn06,Badziag08},
experimental adaptive witnesses~\cite{Park10,Laskowski12}, and
others (see, e.g., Refs.~\cite{Huber10,Jungnitsch11,Rudnicki11,
Rudnicki12,Rudnicki14}.

Some of such studies of entanglement witnesses were focused on the
quantitative description of {entanglement (see, e.g., recent
Refs.~\cite{Osterloh10,Puentes10,Jurkowski10,Liang11,Silvi11,
Lee12,Osterloh12,Ryu12,Zhang13,Rafsanjani13} and a
review~\cite{Guhne09} for older references). For example, a lower
bound on a generic entanglement measure can be derived from the
mean value of entanglement witnesses based on the Legendre
transform~\cite{Guhne07,Eisert07,Guhne08}. The estimation of the
concurrence and/or negativity from entanglement witnesses was
studied in, e.g.,
Refs.~\cite{Brandao05,Breuer06,Audenaert06,Mintert07a,Datta07,Genoni08,Jurkowski10,Chen12}.
In particular, the violation of a Bell inequality, which is also
an entanglement witness,} can be used to estimate the
concurrence~\cite{Verstraete02}, the
negativity~\cite{Bartkiewicz13chsh}, or the relative entropy of
entanglement (REE)~\cite{Horst13}.  A related problem is the
estimation of one entanglement measure from another entanglement
measure, e.g., the concurrence from
negativity~\cite{Verstraete01,Miran04,Datta07} or the
REE~\cite{Miran04b}, the negativity from the REE~\cite{Miran08},
or vice versa.

These approaches based on entanglement witnesses are useful and
efficient, but still their usage is limited, because some
information about the state should be known prior to its
measurement.

{In this paper, we study a universal entanglement witness
(UWE), which can be used as a sufficient and necessary test of the
entanglement of a two-qubit system. The UWE is defined as the
determinant of a partially-transformed density matrix,
$\det\hat\rho^\Gamma$~\cite{Horodecki09}. This witness can be
given as a function of the moments $\Pi_n =
\tr[(\hat\rho^\Gamma)^n]$~\cite{Augusiak08}, which are directly
measurable, as recently described in Ref.~\cite{Bartkiewicz14drg}
for a linear-optical setup. The proposed setup is based on the
experimental methods described and referenced
in Refs.~\cite{Bartkiewicz13discord,Bula13QND,Bartkiewicz13fidelity}.}

{Here we address the problem of applying the UWE to quantify
two-qubit entanglement. Namely, the question is whether the UWE
(or more precisely, its negative expectation value) can be
considered a good entanglement measure. We will show that this is
not the case for arbitrary two-qubit states. However, we will
identify various classes of states for which the UWE is indeed a
good entanglement measure. Moreover, as one of the main results of
this paper, we will demonstrate that the negativity can be given
as a function of the experimentally-accessible moments $\Pi_n$. We
will also discuss how the imperfect measurements of $\Pi_n$
deteriorate the estimation of the negativity.}

This paper is organized as follows: In Sec.~II, the UWE is defined
via experimentally-accessible moments $\Pi_n$ of a given
partially-transposed density matrix. We also show the relation
between the entanglement witness and Makhlin's invariants.  In
Sec.~III, we present one of the main results of our paper, which
is the explicit formula of the negativity as a function of the
experimentally-accessible moments $\Pi_n$. In Sec.~IV, we
demonstrate when the UWE can be considered a useful entanglement
measure. In Sec.~V, we show the relation between the entanglement
witness and the concurrence for an important class of two-qubit
states, namely, the $X$-states. Our results are summarized in
Sec.~VI, as well as Tables~I and~II.

%------------------------------------------------------------------
\section{Universal entanglement witness and Makhlin invariants}

Arguably, the simplest two-qubit separability condition {(Peres-Horodecki 
separability criterion~\cite{Peres96,Horodecki96})} 
can be formulated as follows~\cite{Horodecki09}: 
A two-qubit state $\hat\rho$ is entangled if and only if $\det\hat\rho^\Gamma < 0$,
where $\hat\rho^\Gamma$ is the  partially-transposed (marked by
$\Gamma$) matrix $\hat\rho$. This theorem can be easily shown by
recalling that the partially-transposed matrix of an arbitrary
entangled two-qubit state has full rank and has exactly one
negative eigenvalue. Thus, one can introduce the UWE $\Wuniv$ for
a two-qubit state $\hat\rho$ defined as an operator for which the
expected value is equal to $\det\hat\rho^\Gamma$. This witness can
also be given in terms of the experimentally-accessible moments
$\Pi_n = \tr[(\hat\rho^\Gamma)^n]$ as follows~\cite{Augusiak08}:
\begin{eqnarray}
W &\equiv& \det\hat\rho^\Gamma  =\<\Wuniv\>:=
\tr\big(\Wuniv\hat\rho^{\otimes 4}\big) \nonumber \\
&=&\tfrac{1}{24}(1-6\Pi_4 + 8\Pi_3 + 3\Pi_2^2 - 6\Pi_2).
\label{Witness}
\end{eqnarray}
For convenience, we call the UWE not only the observable $\Wuniv$
but also its expectation value $W$. (This convention is also used
in, e.g., Ref.~\cite{Bartkowiak11} and references therein). In
order to directly measure the UWE one could perform joint
measurements on the four copies $\hat\rho^{\otimes 4}$ of a
two-qubit state $\hat\rho$~\cite{Augusiak08}. A direct and
efficient method for the measurement of $\<\Wuniv\>$ has been
recently proposed for polarization qubits in a linear optical
setup~\cite{Bartkiewicz14drg}. The witness $\Wuniv$, contrary to a
typical entanglement witness, is invariant under local unitary
operations, which follows from the invariance of the moments of
the partially-transposed density matrix that forms the witness.
This invariance is a key requirement of a good entanglement
measure (see Sec. IV and,  e.g., Ref.~\cite{Mintert07}).

It is worth stressing that the moments $\Pi_n$, in
Eq.~(\ref{Witness}) for $n=2,3,4$, are not independent. To show
the connection between these moments, let us analyze them in terms
of the correlation matrix $\hat{\beta}$, with elements $\beta_{ij}
= \mathrm{tr}[(\hat\sigma_i \otimes\hat\sigma_j)\hat\rho]$, and
the Bloch vectors $\mathbf{s}$ and $\mathbf{p}$, with elements
$s_i = \mathrm{tr}[({\hat\sigma}_i\otimes{\hat\sigma}_0)\hat\rho]$
and $p_j =
\mathrm{tr}[(\hat\sigma_0\otimes{\hat\sigma}_j)\hat\rho]$. The
matrices $\hat\sigma_i$ for $i=1,\,2,\,3$ are the Pauli matrices,
and $\hat\sigma_0$ is the single-qubit identity matrix. As shown
in Ref.~\cite{Bartkiewicz14drg}, we can write the first four
moments as
%\begin{subequations}
\begin{eqnarray}
\Pi_1 &=& 1,\nonumber \\
4\Pi_2 &=&   1 + x_1,\nonumber\\
16\Pi_3 &=& 1 + 3x_1 + 6x_2,\nonumber\\
64\Pi_4 &=& 1 + 6x_1+  24x_2 + x_1^2 + 2x_3 +4x_4,
\end{eqnarray}
%\end{subequations}
where %\begin{subequations}
\begin{eqnarray}
x_1 &=&  I_2 + I_4 + I_7,\quad
x_2 = I_1 + I_{12},\nonumber\\
x_3 &=& I_2^2 - I_3, \quad\quad\quad x_4 = I_5 + I_8 + I_{14}
+ I_4I_7,
\end{eqnarray}
%\end{subequations}
are the functions of 9 out of the 18 Makhlin invariants
\cite{Makhlin00}, i.e., $I_1=\det\hat\beta$,
$I_2=\mathrm{tr}(\hat\beta^T\hat\beta)$,
$I_3=\mathrm{tr}(\hat\beta^T\hat\beta)^2$, $I_4=\mathbf{s}^2$,
$I_5=[\mathbf{s}\hat\beta]^2$, $I_7 = \mathbf{p}^2$, $I_8 =
[\hat\beta\mathbf{p}]^2$, $I_{12} =
\mathbf{s}\hat\beta\mathbf{p}$, and $I_{14} =
e_{ijk}e_{lmn}s_ip_l\beta_{jm}\beta_{kn}$, where $e_{ijk}$ is the
Levi-Civita symbol. The invariants are also a subset of 21
fundamental and independent two-qubit invariants described by King
and Welsh in Ref.~\cite{King06}. This demonstrates explicitly
that, in general, in order to measure the UWE one needs to measure
these nine fundamental physical quantities (invariants). Any
function of invariants is also an invariant. We can, therefore,
introduce the following six independent invariants that need to be
measured to estimate the values of moments $\Pi_n$ for $n=2,3,4$.
These invariants are
\begin{eqnarray}
y_1  &=& I_2,\quad y_2  = I_4, \quad y_3  =  I_7, \quad
y_4  = I_1 + I_{12,}\nonumber \\
y_5  &=&  I_5 + I_8 + I_{14},\quad y_6 =  I_3.
\end{eqnarray}
This means that in order to quantify entanglement via the UWE, one
needs to measure exactly six instead of nine independent
quantities. The number of necessary measurements is by 10 smaller
than the number of measurements needed for a full quantum-state
tomography. We can conjecture that this is the minimum number of
independent measurements needed for estimating the entanglement of
an arbitrary two-qubit state.

%------------------------------------------------------------------
\section{Negativity via moments of $\hat\rho^\Gamma$}

{In order to quantify the Peres-Horodecki separability
criterion~\cite{Peres96,Horodecki96}, \.Zyczkowski {\it et
al.}~\cite{Zyczkowski98} introduced a parameter later referred to
as negativity. Subsequently, Vidal and Werner~\cite{Vidal02}
proved that the negativity is an entanglement monotone, so can be
used as an entanglement measure. The negativity has an operational
meaning as the entanglement cost under operations preserving the
positivity of partial transpose
(PPT)~\cite{Audenaert03,Ishizaka04}. It can also be
used as an estimator of entangled dimensions, i.e., to estimate
the number of entangled degrees of freedom of two
subsystems~\cite{Eltschka13}.} The negativity of a 
two-qubit state $\hat\rho$ is usually defined as
\begin{equation}\label{eq:Neg}
N = 2\max\{0,-\min[{\rm eig}(\hat\rho^\Gamma)]\},
\end{equation}
in terms of the minimum (negative) eigenvalue $\lambda \equiv
-\mu=\min[{\rm eig}(\hat\rho^\Gamma)]$ ($\mu>0$) of the
partially-transposed density matrix $\hat\rho^\Gamma$. The task of
finding $\lambda$ is usually not easy because the operation of
partial transposition is not physical, so this operator can only
be implemented approximately.

There is another approach based on measuring moments $\Pi_n =
\tr(\hat\rho^\Gamma)^n$ of the partially-transposed matrix
$\hat\rho^\Gamma$ of a two-qubit state $\hat\rho$.  It has
recently been shown in Ref.~\cite{Bartkiewicz14drg} that all
four first moments $\Pi_n$ can be measured directly using at most
four copies of the investigated two-qubit state. This was shown on
the example of the measurement of a two-photon polarization state
by using a linear-optical setup. We note that this approach of
Ref.~\cite{Bartkiewicz14drg} can be generalized to other
implementations of qubits and various setups.

The first two moments $\Pi_n$ are equivalent to the trace and
purity of $\hat\rho$, i.e., $\Pi_1=1$ and $\Pi_2=p$ respectively.
An efficient method for measuring the purity of an arbitrary
polarization state of two photons has been proposed recently in
Ref.~\cite{Bartkiewicz13fidelity}. The higher-order moments
$\Pi_{3}$ and $\Pi_{4}$ can be measured as described in
Ref.~\cite{Bartkiewicz14drg}. Let us also mention that, as long as
there is some entanglement, $\hat\rho^\Gamma$ has four nonzero
eigenvalues, among which only one is negative and equals to
$\lambda$. This property holds for an arbitrary two-qubit
state~\cite{Rana13}.

Let us  derive an expression for the negativity in terms of the
experimentally-accessible moments $\Pi_n$ for $n=1,2,3$. The
principal invariants of the partially-transposed density matrix
read
\begin{eqnarray}
J_1 &=& \Pi_1= \tr\hat\rho^\Gamma= \lambda_1 + \lambda_2 + \lambda_3 -\mu = 1,\\
J_2 &=& \tfrac{1}{2}(\Pi^2_1-\Pi_2) =\tfrac{1}{2}[(\tr\hat\rho^\Gamma)^2 -\tr(\hat\rho^\Gamma)^2] \nonumber\\
&=& \lambda_1\lambda_2 + \lambda_2\lambda_3 + \lambda_3\lambda_1 -\mu(1+\mu),\\
J_3 &=& \det\hat\rho^\Gamma =- \lambda_1\lambda_2\lambda_3\mu,
\end{eqnarray}
{where $\lambda_n$ for $n=1,\,2,\,3$ are the positive eigenvalues $\rho^\Gamma$ 
and $\mu$ is the module of the negative eigenvalue.}
After simple algebraic manipulations we  derive
\begin{eqnarray}
\sum^3_{n=1}[
-\lambda_n\mu^3 +\tfrac{ 1}{2}\Pi_2\lambda_n\mu + (\lambda_n^2 - \lambda_n)\mu^2 &&\nonumber\\
- \frac{1}{2}(2\lambda_n^3 - 2\lambda_n^2 + \lambda_n)\mu -
\det\hat\rho^\Gamma] &=& 0.
\end{eqnarray}
This sum can be directly calculated using the definition of  moments $\Pi_n$.
As a result we obtain the following expression
\begin{eqnarray}
 -3\mu^4 - 3\mu^3 + \tfrac{3}{2}(\mu^2 + \mu)\Pi_2 -\Pi_3\mu - \tfrac{3}{2}\mu^2&&\nonumber\\ - 3\det\hat\rho^\Gamma -\tfrac{1}{2}\mu &=&0.\label{eq:mu}
\end{eqnarray}
Equation~(\ref{eq:mu}) has an important
consequence, i.e., we can calculate the negativity $N$ after
measuring $\Pi_n$ for $n=1,\,2,\,3,\,4$. As discussed in
Ref.~\cite{Bartkiewicz14drg} measuring these four
experimentally-accessible moments can be done more efficiently
than performing full quantum state tomography.
Equation~(\ref{eq:mu}) is a fourth degree polynomial in $N=2\mu$, which
after simplification reads
\begin{eqnarray}
48\det\hat\rho^\Gamma + 3N^4 + 6N^3 - 6N^2(\Pi_2 - 1) &&\nonumber\\
- 4N(3\Pi_2 - 2\Pi_3 - 1)  &=& 0.\label{eq:N}
\end{eqnarray}
In our opinion this is one of the main results of this work. We
can be sure that it has solutions if $\det\hat\rho^\Gamma<0$
(i.e., when the state $\hat\rho$ is entangled). The solutions can
be found analytically by applying the well-known  Ferrari and
Cardano formulas. Equation~(\ref{eq:N}) has four solutions,
however there is only one real  solution where $N>0$. Therefore,
the value of the negativity is uniquely defined by
Eq.~(\ref{eq:N}). We do not give these solutions explicitly, as
they are lengthy and can be easily obtained by using a computer
algebra system.

\begin{figure}
\includegraphics[width=8.5cm]{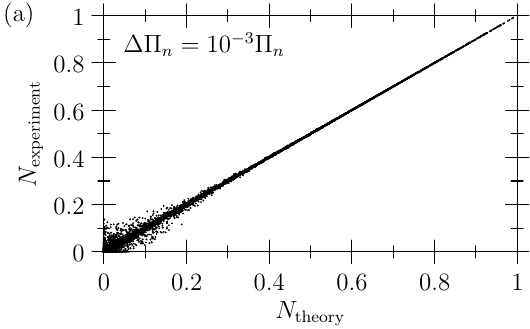}
\includegraphics[width=8.5cm]{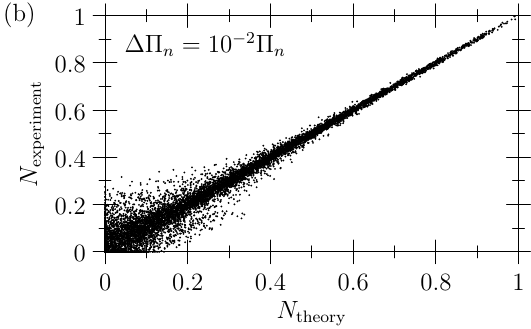}
\caption{\label{fig:errors} The relation between the theoretical
precise values, $N_{\mathrm{theory}}$,   and ``experimental''
noisy values, $N_{\mathrm{experiment}}$, of the negativity for
$10^4$ density matrices randomly generated in a Monte-Carlo
simulation. For each density matrix, all these moments $\Pi_n$
were calculated and then a random noise $\delta \Pi_n\in [0,\Delta \Pi_n]$ was added.
The maximal noise $\Delta\Pi_n$ was set to (a) $10^{-3}\Pi_n$ and
(b) $10^{-2}\Pi_n$. This figure demonstrates that the value of the
experimentally-estimated negativity is very sensitive to noise,
especially if this value is approaching $0$.}
\end{figure}

Unfortunately, the value of  the negativity calculated from
Eq.~(\ref{eq:N}) is very sensitive to the uncertainty of measuring
$\Pi_n$. This can be observed in Fig.~\ref{fig:errors}, where the
relation between the theoretical and the experimentally measured
values of the negativity is depicted for several values of the
maximal relative uncertainties in estimating $\Pi_n$.
Figure~\ref{fig:errors}(b) suggests that if the relative error is
close to $1\%$, the noise level starts  to be too high for
estimating the negativity with a reasonable precision in its
entire range. For uncertainty levels $\gtrsim 10\%$, the
measurement method is not reliable for any value of $N$. Note
that, in all cases, the level of the uncertainty in estimating the
negativity is the largest for the values of $N\approx 0$.

%-----------------------------------------------------------------
\section{Universal entanglement witness as an entanglement measure}

\begin{figure}
\includegraphics[width=8.5cm]{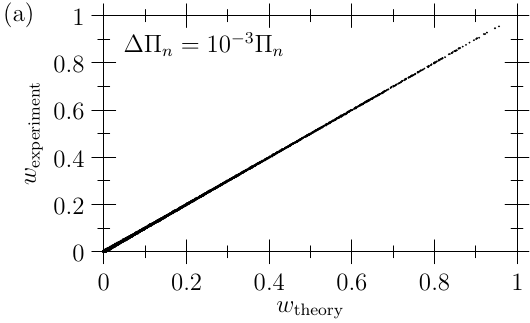}
\includegraphics[width=8.5cm]{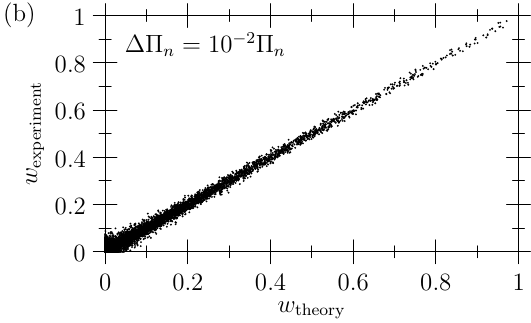}
\caption{\label{fig:errorsW} Same as in Fig.~1 but for the
relation between the theoretical ($w_{\mathrm{theory}}$) and
experimental ($w_{\mathrm{experiment}}$) values of the UWE $w$,
given in Eq.~(\ref{eq:w}). This figure demonstrates how the value
of the experimentally-estimated witness is sensitive to noise. }
\end{figure}

As shown in Fig.~\ref{fig:errorsW}, measuring $\Wuniv$ is less
prone to noise than estimating the negativity.

It is convenient to use the rescaled value of $\<\Wuniv\>$ defined
as
\begin{equation}\label{eq:w}
w:=\max\big[0,-16\<\Wuniv\>\big].
\end{equation}
Now we explicitly describe that the UWE $w$ satisfies the
following standard criteria for a good entanglement measure (as
listed in, e.g., Ref.~\cite{Horodecki09}):

\begin{description}

\item[C1.] The inequalities hold $0\le w(\hat\rho)\le 1$, where
$w(\hat\rho) = 0$ for any unentangled state and $w(\hat\rho) = 1$
for the Bell states.

\item[C2.]
Any local unitary transformations of the form $U_A\otimes U_B$ do
not change $w(\hat\rho)$ for any state $\hat\rho$.

\item[C3.]
An additional property: The witness $w(|\psi\>)$ is simply related
to the entropy of entanglement for any pure state $|\psi\>$, i.e.,
by a relation corresponding to the Wootters formula for the entanglement of
formation~\cite{Wootters98},
\begin{eqnarray}
E_{F}(w) = h\left(\frac{1}{2}\Big[1+\sqrt{1-\sqrt{w}}\Big]\right),
\label{Wootters}
\end{eqnarray}
where $h(y)=-y\log _{2}y-(1-y)\log _{2}(1-y)$ is binary entropy.
This property follows from the observation that
\begin{equation}
  w(|\psi\>)=N^4(|\psi\>)=C^4(|\psi\>)
\end{equation}
{where $w(|\psi\>),\,N(|\psi\>)$, and concurrence $C(|\psi\>)$
are defined by Eqs.~(\ref{eq:w}), (\ref{eq:Neg}), and (\ref{eq:C}), respectively, where
$\rho$ is replaced by $|\psi\>\<\psi|$. This corresponds to case~1 in Table~I.}

\end{description}
Unfortunately, in general, the following two important properties
do not hold for the witness $w(\hat\rho)$.
\begin{description}
\item[C4.]
A good entanglement measure $E(\hat\rho)$ should not increase for
any state $\hat\rho$ and any local operations with classical
communication (LOCC). This property can be violated for
$w(\hat\rho)$ as shown in Appendix~B.

\item[C5.] A good entanglement measure $E(\hat\rho)$ should be convex under
discarding information, i.e., $\sum_i p_i E(\hat\rho_i) \ge
E(\sum_i p_i \hat\rho_i)$. In other words, one cannot increase
$E(\hat\rho)$ by mixing states $\hat\rho_i$. An example of the
violation of this property for $w(\hat\rho)$ is given in
Appendix~C.

\end{description}

Property C.2 follows from the fact that the UWE can be expressed
as a function of local polynomial invariants~\cite{Makhlin00}. For
pure states, the UWE is equivalent to the so-called $G$
concurrence~\cite{Augusiak08,Sinolecka12}, which is a monotone
under LOCC (C.3). Thus, even if the properties C.4 and C.5 are not
satisfied in general, the witness $w$ for two-qubit states is a
useful parameter for quantifying entanglement.

Moreover, the UWE $w$ provides tight upper and lower bounds for
the negativity $N(\hat\rho)$ of an arbitrary two-qubit state
$\hat\rho$~\cite{Augusiak08}:
\begin{equation}\label{eq:bounds}
f(w) \leq N  \leq \sqrt[4]{w},
\end{equation}
where $f(w)$ is the inverse of the polynomial $w(N)=N(N+2)^{3}/27$
on the interval $N\in [0,1]$. Explicitly, the lower bound is given
by~\cite{Bartkiewicz14drg},
\begin{equation}
\label{eq:function_f}
f(w) = \tfrac{1}{2}(-3 + \sqrt{z} + \sqrt{3 - z + \tfrac{2}{\sqrt{z}}}),
\end{equation}
where $z = 1 - y + x$, $y = 36 w/x$, and
\begin{equation}
x = 3 \sqrt[3]{2 \sqrt{w^2 (16 w+1)}-2 w}.
\end{equation}
We show in Table~\ref{tab:xstates} that the states saturating the
upper and lower bounds are pure (case~$1$) and Werner's states
(case~$8$)~\cite{Werner89}, respectively.

The boundary states can be found in the set of the so-called
$X$~states. These states can be simply manipulated~\cite{Horst13}
and are universal in the sense that an arbitrary two-qubit state
can be converted, by a unitary transformation, into its $X$-state
counterpart~\cite{Mendonca14}. Moreover, the $X$~states appear as
solutions in many simple physical models in, e.g., the $XYZ$
Heisenberg model~\cite{Rigolin04,Chen10} or decaying entangled
qubits coupled to a common reservoir exhibiting the effects of
sudden death~\cite{Yu04} and rebirth~\cite{Ficek08} of
entanglement. The name of these states becomes clear when its
density matrix $\hat\rho$ is given explicitly in the standard
computational basis, i.e.,
\begin{equation}
\hat\rho = \left(
\begin{array}{cc|cc}
a & 0 & 0 & b\\
0 & c & d & 0\\
\hline
0 & d^*& e & 0\\
b^* & 0 & 0 & f
\end{array}
\right).\label{eq:X}
\end{equation}
The partial transpose with respect to the second subsystem of
two-qubit density matrix $\hat\rho$ reads
\begin{equation}
\hat\rho^\Gamma = \left(
\begin{array}{cc|cc}
a & 0 & 0 & d\\
0 & c & b & 0\\
\hline
0 & b^*& e & 0\\
d^* & 0 & 0 & f
\end{array}
\right).
\end{equation}
Now, it follows from the Laplace expansion that the UWE for the
$X$~states can be given as a product of determinants,
\begin{eqnarray}
W &=& \det\left(
\begin{array}{cc}
a & d\\
d^* & f
\end{array} \right)
\det\left(\begin{array}{cc}
c & b\\
b^* & e
\end{array}\right) \nonumber\\
&=& \det\left(
\begin{array}{cc}
a & |d|\\
|d| & f
\end{array} \right)
\det\left(\begin{array}{cc}
c & |b|\\
|b| & e
\end{array}\right).
\end{eqnarray}
This is a four-dimensional volume (a product of two areas). We can
expand it further to obtain
\begin{equation}
W  =
[(|d|-\sqrt{af})(|d|+\sqrt{af})][(|b|-\sqrt{ce})(|b|+\sqrt{ce})].
\label{DetRhoGammaX}
\end{equation}
This corresponds to the volume of a four-dimensional box. Note
that the length of its longest negative edge (the longest edge of
negative orientation) corresponds to the negativity. However, the
expression for the negativity is not simple because it requires
finding the smallest eigenvalue of $\hat\rho^\Gamma$, i.e.,
factorizing $\det\hat\rho^\Gamma$ in another way.

%------------------------------------------------------------------
\section{Universal entanglement witness and concurrence}

Remarkably, the largest negative factor in the expression for the
UWE, given by Eq.~(\ref{DetRhoGammaX}), for the $X$~states,
corresponds to another popular entanglement measure. Namely, the
Wootters concurrence~\cite{Wootters98}:
\begin{equation}\label{eq:C}
C({\hat\rho})=\max \Big(0,2\lambda_{\max}-\sum_j\lambda_j\Big),
\end{equation}
where $\lambda^2 _{j} = \mathrm{eig}[{\hat\rho }({\hat\sigma
}_{2}\otimes {\hat\sigma }_{2}){\hat\rho}^{\ast }({ \hat\rho
}_{2}\otimes {\hat\sigma }_{2})]_j$ and
$\lambda_{\max}=\max_j\lambda_j$. The witness $W$ can be
interpreted as a geometric mean of all the lengths in
Eq.~(\ref{DetRhoGammaX}). Thus, the UWE is not a good measure of
entanglement, because it underestimates the available
entanglement. However, the UWE can be used as a measure of
entanglement if  all the edges have the same length and the volume
is negative. {Additional information about the relation 
between the UWE and concurrence for two-qubit states 
can be found in Ref.~\cite{Demianowicz11}.}

Let us note that there are some constrains on the matrix elements
of the $X$~states, e.g., the trace of the partially-transposed
matrix equals $1$. From this observation follows that $a + c + e +
f = 1.$ Other constrains are imposed by the fact that $\hat\rho$
is positive semidefinite, i.e., $|d| \leq \sqrt{ce}$ and  $|b|
\leq \sqrt{af}$. By recalling some properties of density matrices,
we can deduce that the UWE is a monotonic function of a proper
entanglement measure, i.e., the concurrence. The concurrence is
given by the following simple expression for
$X$~states~\cite{Yu07}
\begin{equation}
C = 2\max (0,|d|-\sqrt{af},|b|-\sqrt{ce}).
\end{equation}
One can see that the  UWE is related to both the negativity and
concurrence for the whole class of the $X$~states. For some
subclasses of the $X$~states, the negativity and concurrence are
equivalent. This happens for pure states, rank-2 Bell-diagonal
states, phase-damped states, Bell states with isotropic noise 
(i.e., the Werner~\cite{Werner89} and Werner-like~\cite{Miran04a} states)
(see cases~$1,2,3$, and $8$ in
Table~\ref{tab:xstates}, respectively). For the amplitude-damped
states (case~$4$ in Table~\ref{tab:xstates}) with the damping
parameter $p = 1-f$, the relation is also simple as $C = \sqrt{N^2
+2fN}$~\cite{Horst13}, although it also involves the damping
parameter $p$.

\begin{figure}
\centerline{\includegraphics[width=7.5cm]{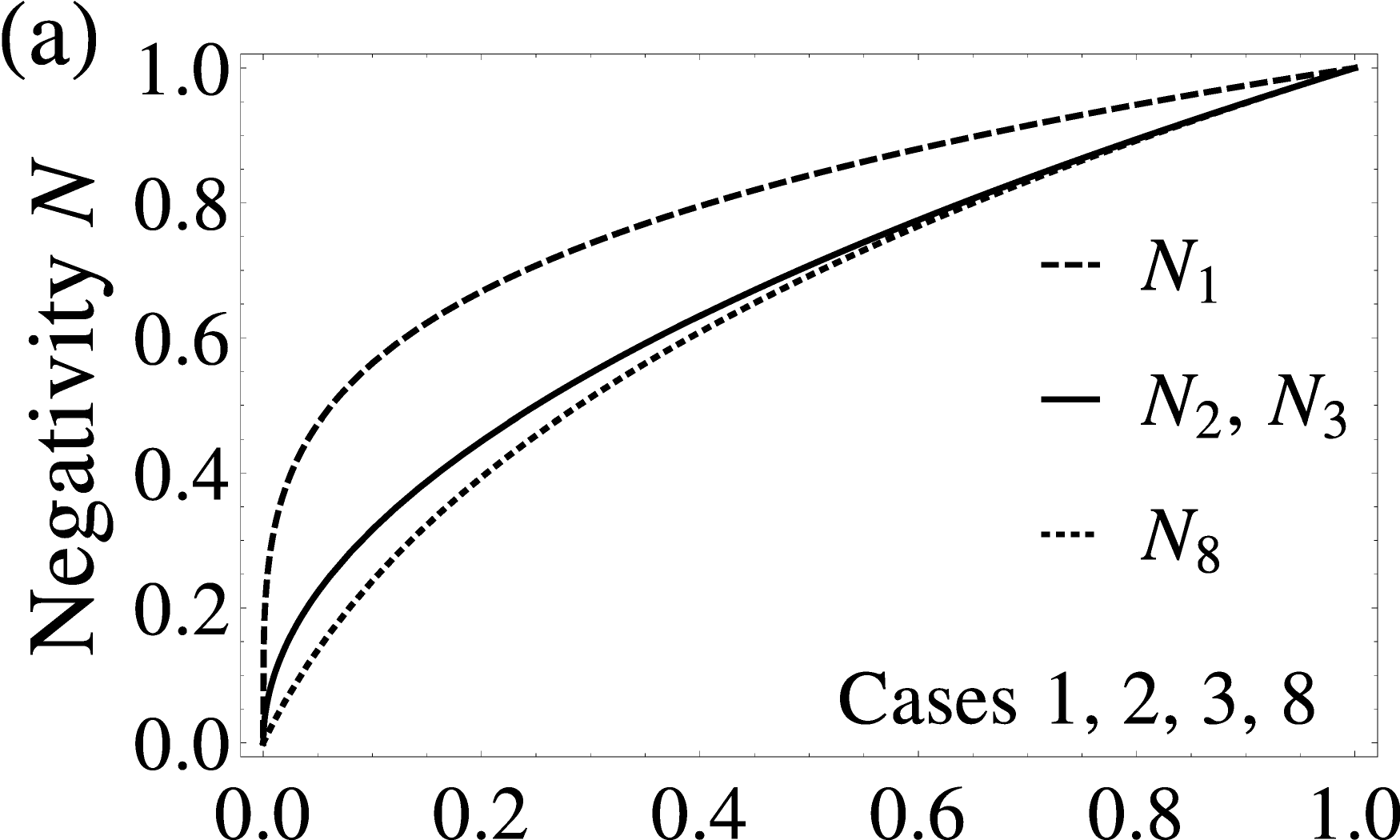}}
\centerline{\includegraphics[width=7.5cm]{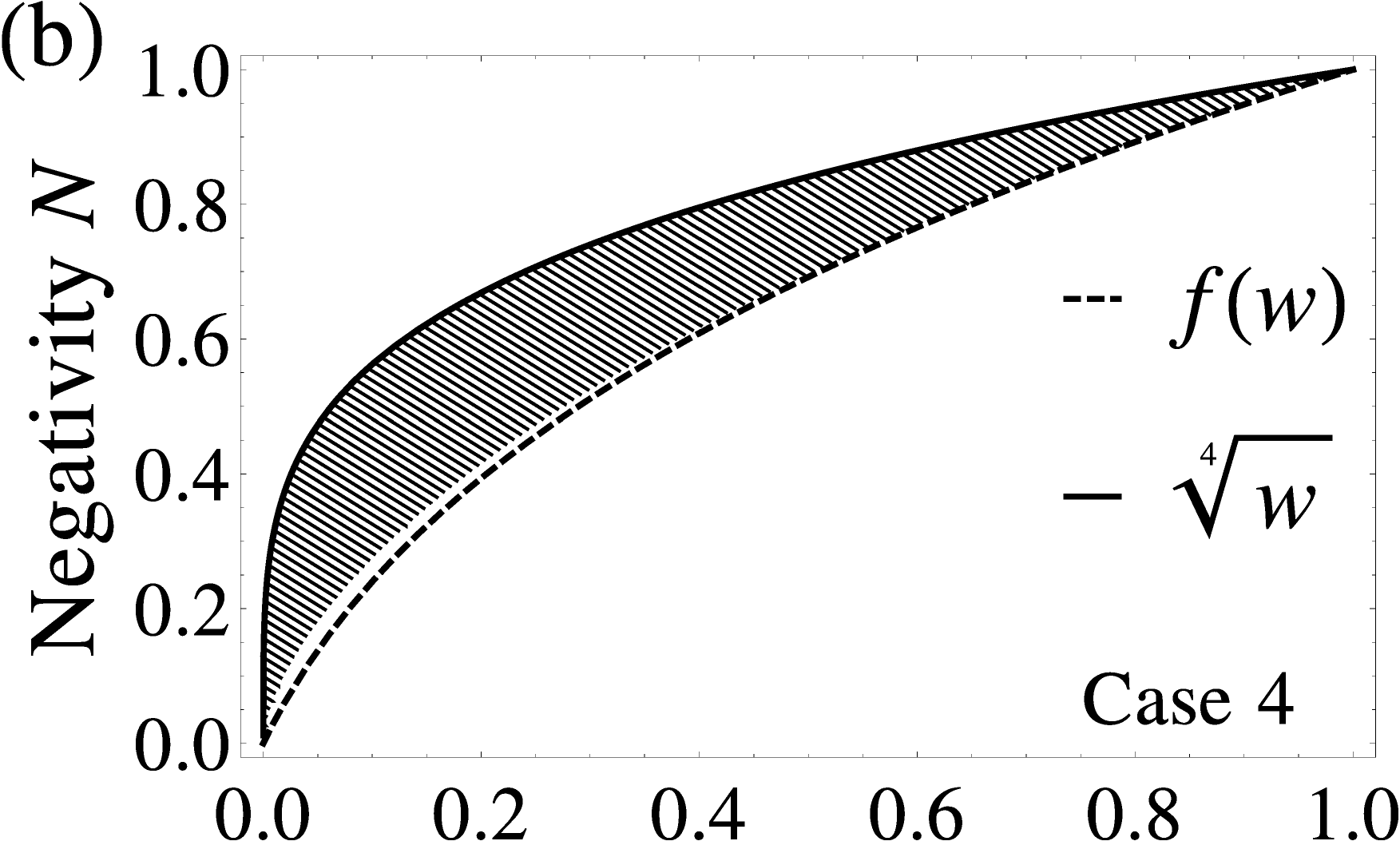}}
\centerline{\includegraphics[width=7.5cm]{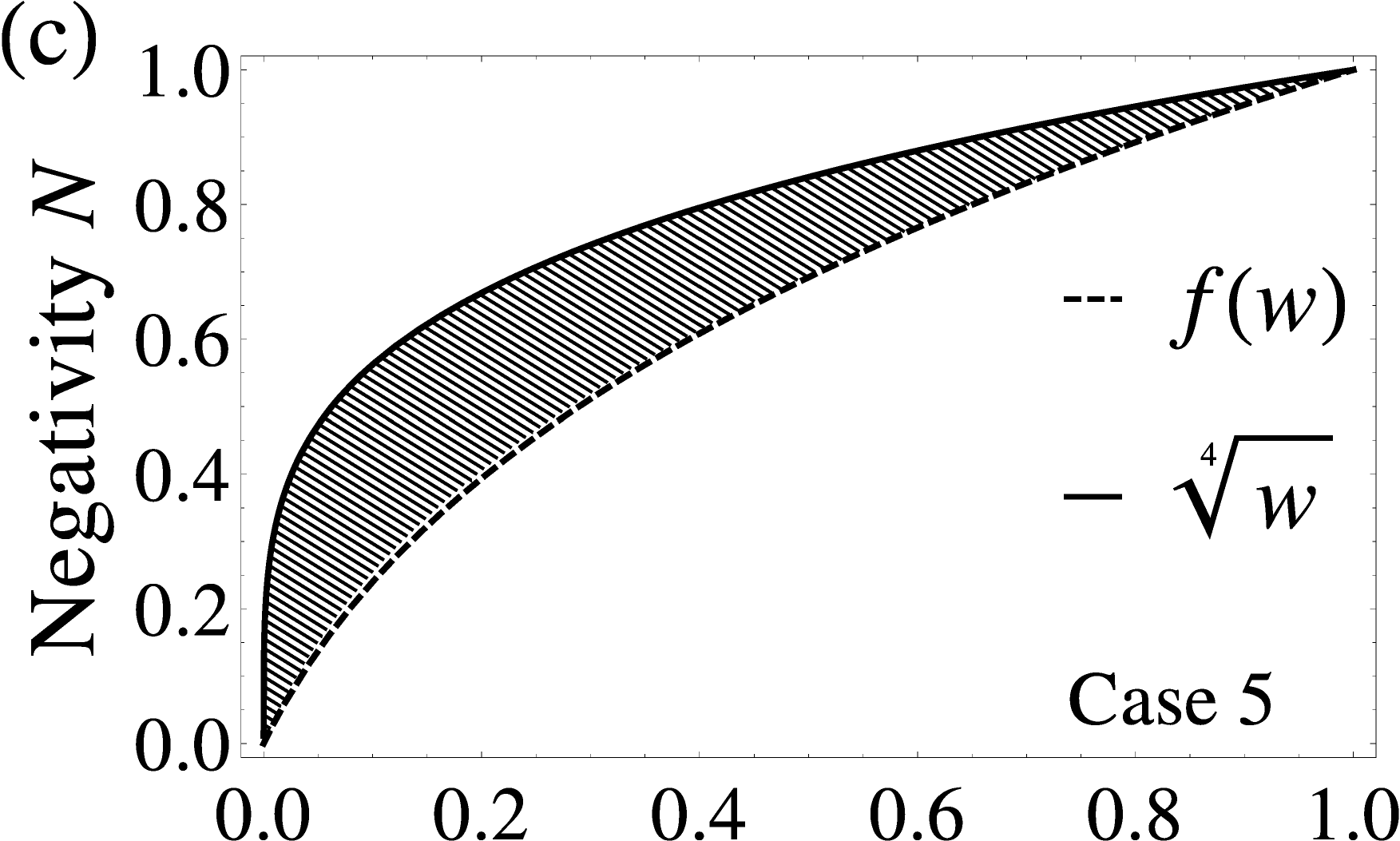}}
\centerline{\includegraphics[width=7.5cm]{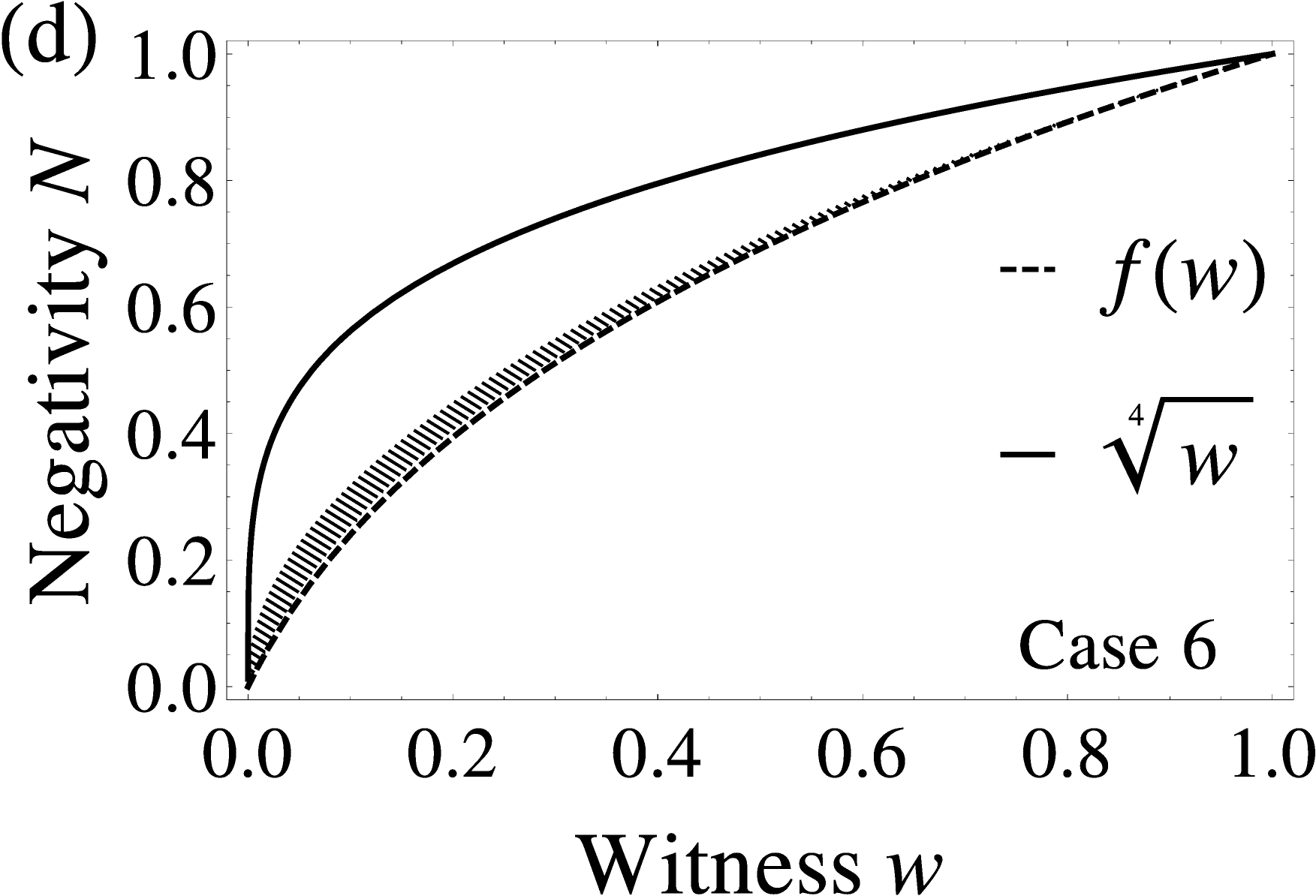}}

\caption{\label{fig:Xstates} Relation between the negativity $N$
and the universal entanglement witness $w$ for various states as
defined in Table~\ref{tab:xstates}. In panel (a) we demonstrate
the relations $N_n(w)$ for states, given in the $n$th case in
Table~\ref{tab:xstates} (b),(c), and (d). The shaded
areas depict the relation $N_n(w)$ for two-parameter states given
in the cases for $n=4,\, 5,\,6$, respectively. The covered area
lies between the dashed curve corresponding to the lower bound
$f(w)$, defined in Eq.~(\ref{eq:function_f}), and the solid curve
corresponding to the upper bound $\sqrt[4]{w}$. In all these
panels, the shaded areas do not cover the whole space between the
boundaries. The whole area is covered only in case~7 which, for
brevity, is not presented here. Strictly speaking, the whole area is \emph{not} covered in panel (c).}
\end{figure}

In Table~\ref{tab:xstates}, we present a survey of the selected
subclasses of the $X$~states of various ranks for which the UWE
(or a function of only $\Pi_2$ and $\Pi_3$ can be considered as an
entanglement measure. In each case the UWE is proportional to a
fourth-degree (or lower-degree) polynomial of $N$ or $C$. For the
states given in cases $1,...,4$, and $8$, the witness $W$ is a
good measure of entanglement because it is a function of $N$ with
constant coefficients. The other states depend on an additional
variable. These states include: the degenerate amplitude-damped
states (case~$5$), rank-3 Bell-diagonal states (case~$6$), and
pure states with isotropic noise (case~$7$). For these states, by
measuring $\Wuniv$ does not provide enough information to
determine the entanglement measures. However, for cases 1, 2, 3,
and 8 listed in Table~I, it is possible to determine $C$ and $N$
by measuring solely $\Pi_2$ and $\Pi_3$. The states of the largest
and smallest ranks are the boundary states for $N$ versus $C$. The
results are also visualized in Fig.~\ref{fig:Xstates}.

Note that we focus only on the states that depend on at most three
independent variables. This is because, by allowing more freedom,
we would have to measure all the first four moments of
$\hat\rho^\Gamma$ to estimate the entanglement. This would give us
no benefit with respect to the approach presented in the previous
section. The states presented in Table~\ref{tab:xstates} may
appear rather specific. Note that $X$~states must be described, in
general, by nine parameters (see, e.g., Ref.~\cite{Hedemann13}).
However, these states represent an infinite set of states that can
be generated by local unitary transformations that do not change
the entanglement. In other words, by applying local unitary
operations, we can always obtain the following rank-specific real
$X$~states~\cite{Hedemann13}:
\begin{eqnarray}
\hat\rho_{1} &\equiv& {\phi^+}(\theta_1),\nonumber \\
\hat\rho_{2a} &\equiv& p_1{\phi^+}(\theta_1) + p_3{\psi^+}(\theta_3),\nonumber \\
\hat\rho_{2b} &\equiv& p_1{\phi^+}(\theta_1) + p_2{\phi^-}(\theta_2),\nonumber \\
\hat\rho_{3} &\equiv& p_1{\phi^+}(\theta_1) + p_2{\phi^-}(\theta_2) + p_3{\psi^+}(\theta_3),\nonumber \\
\hat\rho_{4} &\equiv& p_1{\phi^+}(\theta_1) + p_2{\phi^-}(\theta_2) + p_3{\phi^+}(\theta_3) \nonumber\\
&& + p_4{\psi^-}(\theta_4), \label{eq:Xstates}
\end{eqnarray}
which are incoherent mixtures of pure states
\begin{equation}
\phi_{\pm}(\theta) = \left(\begin{array}{cccc}
\cos^2\theta & 0 & 0 & \pm \tfrac{1}{2} \sin{(2\theta)}\\
0 & 0 & 0 & 0\\
0 & 0 & 0 & 0\\
\pm \tfrac{1}{2} \sin{(2\theta)} & 0 & 0 &  \sin^2\theta
\end{array}\right)
\label{Bell1}
\end{equation}
and
\begin{equation}
\psi_{\pm}(\theta) = \left(\begin{array}{cccc}
0 & 0 & 0 & 0\\
0 & \cos^2\theta & \pm \tfrac{1}{2} \sin{(2\theta)} & 0\\
0 & \pm \tfrac{1}{2} \sin{(2\theta)} & \sin^2\theta & 0\\
0 & 0 & 0 &  0
\end{array}\right)
\label{Bell2}
\end{equation}
with weights $p_i > 0$ ($\sum_i p_i = 1$). Note that the states,
given in Eqs.~(\ref{Bell1}) and~(\ref{Bell2}), reduce to the Bell
states for $\theta=\pi/4$. One should be careful not to accidently
reduce the rank of a given state by choosing some specific values
of $\theta$. The relation between the $X$~states from
Table~\ref{tab:xstates} and the states defined in
Eq.~(\ref{eq:Xstates}) is presented in Table~\ref{tab:params}.

\begin{table*}
\caption{\label{tab:xstates} {A survey on the relation between the
concurrence $C$, negativity $N$, and universal entanglement witness $W$ for selected
subclasses of the $X$~states, where $\vec{x}=(a,b,c,d,e)$ and $g_n
= 1-nf$. Note that $C$ and $W$ can be determined, in some cases,}
by measuring only $\Pi_3$ and (or) $\Pi_2$. However, in general,
these entanglement measures can be obtained by measuring all the
four moments of $\hat\rho^\Gamma$. The presented states of various ranks $R$ include: (Case 1) pure 
states, (2) rank-2 Bell-diagonal states, (3) phase-damped Bell states, 
(4) amplitude-damped pure states, (5) degenerate 
amplitude-damped states, (6) rank-3 Bell-diagonal states, (7) pure 
states with multimode noise, and (8) Werner states (rank-4 Bell-diagonal states). The moments $\Pi_4$ for
all these eight subclasses of the $X$~states are given explicitly
in Appendix~\ref{App:A}. }
\begin{ruledtabular}
\begin{tabular}{clccccc}
Case &$\hat\rho_R = \hat\rho(\vec{x})$ & $R$ & $\Pi_2$ & $\Pi_3$
%& $\Pi_4$
& $W$ & $C$\\
\hline

$1$&
$\left\lbrace\begin{array}{l}
a = b = f = 0\\
|d|=\sqrt{ce}
\end{array}\right.$
& $1$
&
$1$
&
$1 - \frac{3}{4}N^2$
%& $\left(1-\frac{N^2}{2}\right)^2$
& $-\frac{N ^4}{16}$
& $2|d|=N$\\
%\hline

$2$&
$\left\lbrace\begin{array}{l}
a = b = f < \frac{1}{2}\\
c = d = e = \frac{g_2}{2}
\end{array}\right.$
& $2a$
&$\frac{1}{2}(N^2 + 1)$
&$\frac{1}{4}$
%& $\left(1-\frac{N^2}{2}\right)^2$
%& $\frac{1}{8}(N^4+ 1)$
& $-\frac{N^2}{16}$
& $|g_4|=N$\\
%\hline

$3$&
$\left\lbrace\begin{array}{l}
a=b=f =0\\
e=c=\frac{1}{2}\\
|d|< \frac{1}{2}\\
\end{array}\right.$
& $2b$
& $\frac{1}{2}(N^2 + 1)$
& $\begin{array}{l}
\frac{1}{4}
\end{array}$
%& $\left(1-\frac{N^2}{2}\right)^2$
%& $\frac{1}{8}(N^4+ 1)$
%& $\frac{1}{8}(N^4+ 1)$
&$-\frac{N^2}{16}$
& $2 | d|=N$\\
%\hline

$4$&
$\left\lbrace\begin{array}{l}
a = b = 0\\
f=1-c-e\\
|d|=\sqrt{ce}
\end{array}\right.$
& $2b$
&
$g_2 + 2f^2$
&
$\begin{array}{l}
1-3\left(1+\frac{C^2}{2}\right)g_1\\
+3g_1^2+\frac{3C^2}{4}
\end{array}$
%& $\left(1-\frac{N^2}{2}\right)^2$
%& $\frac{1}{8}(N^4+ 1)$
%& $\frac{1}{8}(N^4+ 1)$
%& $\begin{array}{l}
%\frac{C^4}{4} - g_2C^2\\
%+ g_1^4 + f^4
%\end{array}$
& $-\frac{C^4}{16}$
& $\begin{array}{l}
2|d|
\\=\sqrt{N^2 +2fN}\end{array}$\\
%\hline

$5$&
$\left\lbrace\begin{array}{l}
b = 0\\
a=f=\frac{1-c-e}{2}\\
|d|=\sqrt{ce}>f
\end{array}\right.$
& $3$
&
$g_4 + 6f^2$
&
$\begin{array}{l}
g_2^3+ 2f^3  \\
- \frac{3}{4}g_4(C+2f)^2
\end{array}$
& $-\frac{C(C+4f)(C+2f)^2}{16}$
& $2|d|-2f$\\
%\hline

%$6$&
%$\begin{array}{l}
%b = 0\\
%a=f=\frac{1-c-e}{2}\\
%|d|=\sqrt{ce}<f
%\end{array}$
%& $3$
%&
%$g_4 + 6f^2$
%&
%$\begin{array}{l}
%g_2^3+ 2f^3  \\
%- \frac{3}{4}g_4(C-2f)^2
%\end{array}$
%&
%$\begin{array}{l}
%[g_2^2 -\frac{1}{2}(C-2f)^2]^2\\
% + 3(Cf-2f^2)^2 + 2f^4
%\end{array}$
%& $-\frac{C(C-4f)(C-2f)^2}{16}$
%& $2f-2|d|$\\
%\hline

$6$&
$
\left\lbrace\begin{array}{lr}
a=b=f<\frac{1}{4} \\
e=c=\frac{g_2}{2}\\
|d| < \sqrt{ce}\\
\end{array}\right.$
& $3$
& $\begin{array}{l}
\frac{C}{8}(3C+2)\\
+ 2|d|^2 + \frac{3}{8}
\end{array}
$
& Eq.~(\ref{A2})
%& $\left(1-\frac{N^2}{2}\right)^2$
%& $\frac{1}{8}(N^4+ 1)$
%& $\frac{1}{8}(N^4+ 1)$
%& $\begin{array}{l}
%\frac{C^4}{4} - g_2C^2\\
%+ g_1^4 + f^4
%\end{array}$
%&
%$\begin{array}{l}
%[g_2^2 -\frac{1}{2}(C+2f)^2]^2\\
% + 3(Cf+2f^2)^2 + 2f^4
%\end{array}$
%& $\begin{array}{l}
%2^{-7}[9C^4 +  4C^3 \\
% + 6(2^4d^2 + 1)C^2  \\
% + 4(48d^2+1)C\\
%  + 2^8d^4+96d^2 + 9]
%\end{array}$
& $\begin{array}{l}
-\frac{C(C^2 +2C +1 -16|d|^2)}{64}
\end{array}$
& $|g_4|$\\
%\hline

$7$&
$\left\lbrace\begin{array}{l}
a = f < |d|\\
b = 0 \\
c = c'+\frac{f}{2}\\
e = e'+\frac{f}{2}\\
|d|=\sqrt{c'e'}-\frac{f}{2}\\
\end{array}\right.$
& $4$
& $g_3^2 + g_3f +\frac{5}{2}f^2$
& Eq.~(\ref{A3})
& Eq.~(\ref{A4})
& $2|d|-2f$\\
%\hline

$8$&
$\left\lbrace\begin{array}{l}
a = f<\frac{1}{6}\\
b = 0\\
c = e = \frac{g_2}{2}\\
|d| = \frac{g_4}{2}
\end{array}\right.$
& $4$
& $\begin{array}{l}
N +{\tfrac{1}{3}(1-N)^2}
\end{array}$
& $\begin{array}{l}
\frac{1}{36} (-4 N^3+3 N^2\\
+6 N+4)
\end{array}$
& $\begin{array}{l}
\frac{1}{16} (\frac{1-N}{3}-1)^3 N
\end{array}$
& $\begin{array}{l}
|g_6|=N
\end{array}$
\end{tabular}
\end{ruledtabular}
\end{table*}

\begin{table}
\caption{\label{tab:params}The relation between the states from
Table~\ref{tab:xstates} and the $X$~states defined in
Eq.~(\ref{eq:Xstates}). The correspondence is valid up to local
unitary transformations on two qubits.}
\begin{ruledtabular}
\begin{tabular}{ccccccccc}
Case & $p_1$ & $p_2$ & $p_3$ & $p_4$ & $\theta_1$ & $\theta_2$ & $\theta_3$ & $\theta_4$\\
\hline
1 & 0 & 0 & 1 & 0 & 0 & 0 & $\mathrm{acos}\sqrt{c}$ & 0 \\
2 & $2f$ & 0 & $g_2$ & 0 & $\frac{\pi}{4}$ & 0 & $\frac{\pi}{4}$ & 0\\
3 & 0 & 0 & $\frac{1}{2}+|d|$ & $\frac{1}{2}-|d|$ & 0 & 0 & $\frac{\pi}{4}$ & $\frac{\pi}{4}$\\
4 & $2f$ & 0 & $g_2$ & 0 & $\frac{\pi}{2}$ & 0 & $\mathrm{acos}\sqrt{\frac{c}{g_2}}$ & 0 \\
5 & $f$ & $f$ &  $g_2$ & 0 & $\frac{\pi}{4}$ & $\frac{\pi}{4}$ &  $\mathrm{acos}\sqrt{\frac{c}{g_2}}$ & 0\\
6 & $2f$ & 0 & $g_2(\frac{1}{2}+|d|)$ & $g_2(\frac{1}{2}-|d|)$ & $\frac{\pi}{4}$ & 0 & $\frac{\pi}{4}$ & $\frac{\pi}{4}$ \\
7 & $f$ & $f$ & $g_3$ & $f$ & $\frac{\pi}{4}$ & $\frac{\pi}{4}$ & $\mathrm{acos}\sqrt{\frac{2c-f}{2g_3}}$ & $\frac{\pi}{4}$\\
8 & $f$ & $f$ & $g_3$ & $f$ & $\frac{\pi}{4}$ & $\frac{\pi}{4}$ & $\frac{\pi}{4}$ & $\frac{\pi}{4}$
\end{tabular}
\end{ruledtabular}
\end{table}

\section{Conclusions}

We have described a direct operational method for determining the
negativity of an arbitrary two-qubit state. We have derived the
method by analyzing the relation between the purity, negativity,
and a universal entanglement witness for two-qubit entanglement.
In particular, we have expressed the negativity as a function of
six invariants which are linear combinations of nine from the
complete set 21 fundamental and independent two-qubit invariants
listed, e.g., in Ref.~\cite{King06}.

We have demonstrated how to measure the negativity of a two-photon
polarization state by measuring three experimentally-accessible
moments $\Pi_n$ of the partially-transposed density matrix of a
two-photon state. We pointed out that this approach can be more
practical than directly estimating the negativity, which is
sensitive even to a low-level noise.

We also discussed the possibility of using the universal
entanglement witness or lower moments of $\hat\rho^\Gamma$ as a
proper entanglement measure for some classes of states. In
particular, we demonstrated their relation to the negativity and
concurrence for the $X$~states.

It is worth noting that the UWE is not necessarily the least-error
sensitive entanglement measure, which can be constructed from the
moments of the partially-transposed density matrix of a given
state. It is possible that a better two-qubit entanglement measure
exists that can be measured as a function of $\Pi_n$ for
$n=2,3,4$.

We hope that these results can pave the way for direct and
efficient methods for measuring two-qubit quantum entanglement.

%------------------------------------------------------------------
\begin{acknowledgments}
We thank Pawe\l{} Horodecki for stimulating discussions. K. L.
gratefully acknowledges the support by the Czech Science
Foundation under the project No. 13-31000P. A.M. is supported by
the Polish National Science Centre under grants
DEC-2011/03/B/ST2/01903 and DEC-2011/02/A/ST2/00305. K.B.
acknowledges the support by the Foundation for Polish Science (START Programme) and
the Polish National Science Centre under grant No.
DEC-2013/11/D/ST2/02638, and the project No. LO1305 of the
Ministry of Education, Youth and Sports of the Czech Republic.
\end{acknowledgments}

\appendix
\section{Moments $\Pi_3$ and $\Pi_4$ for some states in Table~I\label{App:A}}

Here we show explicitly $\det\hat\rho^\Gamma$ and the moments
$\Pi_3$ and $\Pi_4$ of the partially-transposed density matrix
$\hat\rho^\Gamma$ for the selected subclasses of the $X$~states
given in Table~\ref{tab:xstates}. These moments are given as a
function of either the concurrence $C$ or the negativity $N$.

The moments $\Pi^{(n)}_4$ for the $n$th case (subclass) of the
$X$~states analyzed in Table~I are the following:
%\begin{subequations}
\begin{eqnarray}
\Pi_4^{(1)} &=&\left(1-\frac{N^2}{2}\right)^2,\nonumber \\
\Pi_4^{(2)} &=&\Pi_4^{(3)} =\tfrac{1}{8}(N^4+ 1),\nonumber \\
\Pi_4^{(4)} &=&\frac{C^4}{4} - g_2C^2
+ g_1^4 + f^4,\nonumber \\
\Pi_4^{(5)} &=& [g_2^2 -\tfrac{1}{2}(C+2f)^2]^2
 + 3(Cf+2f^2)^2
 + 2f^4,\nonumber \\
\Pi_4^{(6)} &=& 2^{-7}[9C^4 +  4C^3
 + 6(2^4d^2 + 1)C^2  \nonumber\\
&& + 4(48d^2+1)C
  + 2^8d^4+96d^2 + 9],\nonumber \\
\Pi_4^{(7)} &=& -\tfrac{3}{4}C^2g_4 - \tfrac{3}{2}f^3 - 3fg_4C +
\tfrac{63}{4}f^2
- \tfrac{15}{2}f + 1, \nonumber \\
\Pi_4^{(8)} &=& \tfrac{1}{108}(7N^4 + 2N^3 + 6N^2 + 8N + 4).
\label{A1}
\end{eqnarray}
%\end{subequations}
where $g_n = 1-nf$, while $d$ and $f$ are the elements of
$\hat\rho$,  given in Eq.~(\ref{eq:X}). The moment $\Pi_3$ for the
$X$~state in case~6 reads
\begin{equation}
\Pi_3 = \tfrac1{32}[3C(1-C^2 +C +16|d|^2) + {48|d|^2 + 5}].
\label{A2}
\end{equation}
The moment $\Pi_3$ and $\det\hat\rho^\Gamma$ for the $X$~states in
case~7 read
\begin{eqnarray}
\Pi_3 &=&\tfrac{1}{4}C^4 +2fC^3
+\tfrac{3}{4}(5f^2-6f-\tfrac{4}{3})C^2+\tfrac{289}{8}f^4
\nonumber\\
&&+f(f^2-18f+4)C
 -\tfrac{89}{2}f^3
+\tfrac{67}{2}f^2 +g_{10}, \label{A3} \end{eqnarray}
\begin{eqnarray}
\det\hat\rho^\Gamma &=& - \tfrac{1}{16}C^4 - \tfrac{1}{2}fC^3
- \tfrac{1}{16}f(15f + 2)C^2 \nonumber \\
&&- \tfrac{1}{4}f^2(2-f)C. \label{A4}
\end{eqnarray}

\section{Violation of the LOCC condition}

Here we show that the LOCC criterion C4., characterizing a good
entanglement measure, can be violated for the  UWE. Thus, we
analyze the following two-qubit Bell-diagonal state
\begin{equation}
\hat\rho = p\psi_-(\tfrac{\pi}{4}) + (1-p)\phi_+(\tfrac{\pi}{4}),
\end{equation}
for which $w(\hat\rho)$ can increase under some local operations,
as shown explicitly below.

As an example of an LOCC operation, we apply the ``twirling''
operation~\cite{Bennett96}, where  a random SU(2) rotation is
performed on each qubit. This twirling changes $\hat\rho$ into the
Werner state
\begin{eqnarray}
\hat\rho' &=& p\psi_-(\tfrac{\pi}{4}) +
\tfrac13(1-p)[\phi_+(\tfrac{\pi}{4}) + \phi_-(\tfrac{\pi}{4}) +
\psi_+(\tfrac{\pi}{4})]
\nonumber \\
&=& q\psi_-(\tfrac{\pi}{4}) + \tfrac14(1-q) I,
\end{eqnarray}
which is a mixture of the singlet state $\psi_-(\tfrac{\pi}{4}),$
with the weight $q=(4p-1)/3$, and the maximally-mixed state as
given by the four-dimensional identity operator $I$. Consequently,
for $p = (3\sqrt{17} - 7)/8,$ we observe the largest violation of
the LOCC condition for this particular state $\hat\rho$. This is
because, $w(\hat\rho) = 0.11719$ and $w(\hat\rho') = 0.16294$,
hence $w(\hat\rho) < w(\hat\rho')$.

It is worth noting that if these twirling operations are applied
to the concurrence, negativity, or the REE, then property C.4 is
always satisfied. Anyway, the twirling operations can be used to
show that the Werner states determine the lower bounds of the
concurrence for a given value of the
negativity~\cite{Verstraete01}, the REE vs
negativity~\cite{Vedral98,Miran04b}, or the REE vs the Bell
nonlocality~\cite{Horst13}.

\section{Violation of the convexity condition}

Here we show that the convexity criterion C5., which is another
important condition for a good entanglement measure, can also be
violated for the UWE and some states.

Thus, let us consider a mixture $\hat\rho = (\hat\rho_1 +
\hat\rho_2)/2$ of the following two-qubit density matrices
\begin{eqnarray}
\hat\rho_1 &=& \tfrac{1}{2}\left[\phi_+(0) + \psi_+(\tfrac{\pi}{8})\right],\\
\hat\rho_2 &=& \tfrac{1}{2}\left[\phi_+(0) +
\psi_-(\tfrac{5\pi}{8})\right].
\end{eqnarray}
For these states, the convexity condition should imply that
\begin{equation}\label{eq:cvx}
w(\hat\rho) \leq \tfrac{1}{2}w(\hat\rho_1) +
\tfrac{1}{2}w(\hat\rho_2).
\end{equation}
However, the relevant values of the UWE read $w(\hat\rho_1) =
2^{-6}$, $w(\hat\rho_2) = 2^{-6}$, and $w(\hat\rho) = 2^{-5}$. It
is seen that $w(\hat\rho_1) + w(\hat\rho_2) = w(\hat\rho)$. Thus,
the convexity condition (\ref{eq:cvx}) is clearly violated because
$w(\hat\rho) \nleq \tfrac{1}{2}w(\hat\rho)$ for
$w(\hat\rho)=\tfrac{1}{32}$.

%------------------------------------------------------------------

%Unused bibitems
%


\begin{thebibliography}{10}
\bibitem{Schrodinger35}
E. Schr\"odinger, \extra{Discussion of Probability Relations
between Separated Systems,} Proc. Camb. Phil. Soc. {\bf 31}, 555
(1935).


\bibitem{Einstein35}
A. Einstein, N. Podolsky, and B. Rosen, \extra{Can
Quantum-Mechanical Description of Physical Reality Be Considered
Complete?,} Phys. Rev. {\bf 47}, 777 (1935).


\bibitem{BengtssonBook}
I. Bengtsson and K. \.Zyczkowski, \emph{Geometry of Quantum
States} (Cambridge University Press, Cambridge, 2006).


\bibitem{Horodecki09} R. Horodecki, P. Horodecki, M. Horodecki, and K.
Horodecki, \extra{Quantum entanglement,} Rev. Mod. Phys. {\bf 81},
865 (2009).


\bibitem{SchleichBook}
W. P. Schleich and H. Walther, \emph{Elements of Quantum
Information} (Wiley-VCH, Weinheim, 2007).


\bibitem{WernerList}
O. Krueger and R.F. Werner (eds.), \extra{Some Open Problems in
Quantum Information Theory}, e-print quant-ph/0504166.


\bibitem{Guhne09}
O. G\"uhne and G. T\'oth, \extra{Entanglement detection,} Phys.
Rep. {\bf 474}, 1 (2009).


\bibitem{ParisBook}
M.G.A. Paris and J. \v{R}eh\'a\v{c}ek (eds.), \emph{Quantum State
Estimation}, Lecture Notes in Physics, Vol. 649 (Springer, Berlin,
2004).


\bibitem{Miran14}
A. Miranowicz, K. Bartkiewicz, J. Perina Jr., M. Koashi, N. Imoto,
and F. Nori, \extra{Optimal two-qubit tomography based on local
and global measurements}, \pra {\bf 90}, 062123 (2014).


\bibitem{Horodecki96}
M. Horodecki, P. Horodecki, and R. Horodecki, \extra{Separability
of mixed states: necessary and sufficient conditions,} \pla
\textbf{223}, 1 (1996).

\bibitem{Bourennane04}
M. Bourennane \etal, \extra{Experimental Detection of Multipartite
Entanglement using Witness Operators,} \prl {\bf 92}, 087902
(2004).


\bibitem{Haffner05}
H. H\"affner \etal, \extra{Scalable multiparticle entanglement of
trapped ions,} Nature (London) {\bf 438}, 643 (2005).


\bibitem{Leibfried05}
D. Leibfried \etal, \extra{Creation of a six-atom `Schr\"odinger
cat' state,} Nature (London) {\bf 438}, 639 (2005).


\bibitem{Bovino05}
F. A. Bovino, G. Castagnoli, A. Ekert, P. Horodecki, C. Moura
Alves, and A. V. Sergienko, \extra{Direct Measurement of Nonlinear
Properties of Bipartite Quantum States,} \prl {\bf 95}, 240407
(2005).


\bibitem{Horodecki02}
P. Horodecki and A. Ekert, \extra{Method for Direct Detection of
Quantum Entanglement,} \prl {\bf 89}, 127902 (2002).


\bibitem{Horodecki03}
P. Horodecki, \extra{From limits of quantum operations to
multicopy entanglement witnesses and state-spectrum estimation,}
\pra {\bf 68}, 052101 (2003).


\bibitem{Carteret05}
H. A. Carteret, \extra{Noiseless Quantum Circuits for the Peres
Separability Criterion,} \prl {\bf 94}, 040502 (2005);
\extra{Exact interferometers for the concurrence and residual
3-tangle,} quant-ph/0309212.


\bibitem{Aolita06}
L. Aolita and F. Mintert, \extra{Measuring Multipartite
Concurrence with a Single Factorizable Observable,} \prl {\bf 97},
050501 (2006).


\bibitem{Walborn06}
S. P. Walborn, P. H. Souto Ribeiro, L. Davidovich, F. Mintert, and
A. Buchleitner, \extra{Experimental determination of entanglement
with a single measurement,} Nature (London) {\bf 440}, 1022
(2006).
L. Zhou and Y.-B. Sheng, Detection of nonlocal atomic entanglement 
assisted by single photons, \pra {\bf 90}, 024301 (2014). 

\bibitem{Badziag08}
P. Badziag, C. Brukner, W. Laskowski, T. Paterek, and M.
\.{Z}ukowski, \extra{Experimentally Friendly Geometrical Criteria
for Entanglement,} \prl {\bf 100}, 140403 (2008).


\bibitem{Park10}
H. S. Park, S. S. B. Lee, H. Kim, S. K. Choi, and H. S. Sim,
\extra{Construction of an Optimal Witness for Unknown Two-Qubit
Entanglement,} \prl {\bf 105}, 230404 (2010).


\bibitem{Laskowski12}
W. Laskowski, D. Richard, C. Schwemmer, T. Paterek, and H.
Weinfurter, \extra{Experimental Schmidt Decomposition and State
Independent Entanglement Detection,} \prl {\bf 108}, 240501
(2012).


\bibitem{Huber10}
M. Huber, F. Mintert,  A. Gabriel, and B. C. Hiesmayr,
\extra{Detection of High-Dimensional Genuine Multipartite
Entanglement of Mixed States,} \prl {\bf 104},  210501 (2010).


\bibitem{Jungnitsch11}
B. Jungnitsch, T. Moroder, and O. G\"uhne, \extra{Taming
multiparticle entanglement,} \prl {\bf 106}, 190502 (2011).


\bibitem{Rudnicki11} {\L}. Rudnicki, P. Horodecki, and K. \.{Z}yczkowski,
\extra{Collective Uncertainty Entanglement Test,} \prl {\bf 107},
150502 (2011).


\bibitem{Rudnicki12} {\L}. Rudnicki, Z. Pucha\l{}a, P. Horodecki, and K. \.{Z}ycz\-kowski
\extra{Collectibility for mixed quantum states,} \pra {\bf 86},
062329 (2012).


\bibitem{Rudnicki14}
{\L}. Rudnicki, Z. Pucha\l{}a, P. Horodecki, and K.
\.{Z}ycz\-kow\-ski, \extra{Constructive entanglement test from
triangle inequality,} J. Phys. A: Math. Theor. \textbf{47}, 424035 (2014). 

\bibitem{Osterloh10} A. Osterloh and P. Hyllus,
\extra{Estimating multipartite entanglement measures,} \pra {\bf
81}, 022307 (2010).


\bibitem{Puentes10}
G. Puentes, A. Datta, A. Feito, J. Eisert, M. B. Plenio, and I. A.
Walmsley, \extra{Entanglement quantification from incomplete
measurements: Applications using photon-number-resolving weak
homodyne detectors,} New J. Phys. {\bf 12}, 033042 (2010).


\bibitem{Jurkowski10}
J. Jurkowski and D. Chru\'sci\'nski, \extra{Estimating concurrence
via entanglement witnesses,} \pra {\bf 81}, 052308 (2010).


\bibitem{Liang11}
Y.C. Liang, T. V\'ertesi, and N. Brunner,
\extra{Semi-device-independent bounds on entanglement,} \pra {\bf
83} 022108 (2011).


\bibitem{Silvi11}
P. Silvi, F. Taddei, R. Fazio, and V. Giovannetti,
\extra{Quantitative entanglement witnesses of isotropic and Werner
classes via local measurements,} J. Phys. A: Math. Theor. {\bf
44}, 145303 (2011).


\bibitem{Lee12}
S.S. B. Lee and H.S. Sim, \extra{Quantifying mixed-state quantum
entanglement by optimal entanglement witnesses,} \pra {\bf 85}
022325 (2012).


\bibitem{Osterloh12}
A. Osterloh and J. Siewert, \extra{Invariant-based entanglement
monotones as expectation values and their experimental detection,}
\pra {\bf 86} 042302 (2012).


\bibitem{Ryu12}
S. Ryu, S.S.B. Lee, and H.S. Sim, \extra{Minimax optimization of
entanglement witness operator for the quantification of
three-qubit mixed-state entanglement,} \pra {\bf 86}, 042324
(2012).


\bibitem{Zhang13}
C. Zhang, S. Yu, Q. Chen, and C. H. Oh, \extra{Detecting and
Estimating Continuous-Variable Entanglement by Local Orthogonal
Observables,}  \prl {\bf 111}, 190501 (2013).


\bibitem{Rafsanjani13}
S. M. Hashemi Rafsanjani, C. J. Broadbent, and J. H. Eberly,
\extra{Bounding the entanglement of $N$ qubits with only four
measurements,}  \pra {\bf  88}, 062331 (2013).


\bibitem{Guhne07}
O. G{\"u}hne, M. Reimpell, and R.F. Werner, \extra{Estimating
entanglement measures in experiments,} \prl {\bf 98}, 110502
(2007).


\bibitem{Eisert07} J. Eisert, F. Brandao, and K. Audenaert,
\extra{Quantitative entanglement witnesses,} New J. Phys. {\bf 9},
46 (2007).


\bibitem{Guhne08}
O. G{\"u}hne, M. Reimpell, R.F. Werner, \extra{Lower bounds on
entanglement measures from incomplete information,} \pra {\bf 77},
052317 (2008).


\bibitem{Brandao05}
F.G.S.L. Brandao, \extra{Quantifying entanglement with witness
operators,} \pra \textbf{72}, 022310 (2005).


\bibitem{Breuer06}
H.-P. Breuer, \extra{Separability criteria and bounds for
entanglement measures,} J. Phys. A: Math. Gen. \textbf{39}, 11847
(2006).


\bibitem{Audenaert06}
K.M.R. Audenaert and M.B. Plenio, \extra{When are correlations
quantum? Verification and quantification of entanglement by simple
measurements,} New J. Phys. \textbf{8}, 266 (2006).


\bibitem{Mintert07a}
F. Mintert, \extra{Concurrence via entanglement witnesses,} Phys.
Rev. A \textbf{75}, 052302 (2007).


\bibitem{Datta07}
A. Datta, S. Flammia, A. Shaji, and C. Caves, \extra{Constrained
bounds on measures of entanglement,} \pra {\bf 75}, 062117 (2007).


\bibitem{Genoni08}
M. Genoni, P. Giorda, and M. Paris, \extra{Optimal estimation of
entanglement,} \pra {\bf 78}, 032303 (2008).


\bibitem{Chen12}
Z.-H. Chen, Z.-H. Ma, O. G\"uhne, and S. Severini,
\extra{Estimating Entanglement Monotones with a Generalization of
the Wootters Formula,} \prl {\bf 109} 200503 (2012).


\bibitem{Verstraete02}
F. Verstraete and M.M. Wolf, \extra{Entanglement versus Bell
violations and their behavior under local filtering operations,}
\prl \textbf{89}, 170401 (2002).

\bibitem{Bartkiewicz13chsh}
K. Bartkiewicz, K. Lemr, B. Horst, and A. Miranowicz,
\extra{Entanglement estimation from Bell inequality violation,}
\pra {\bf 88}, 052105 (2013).

\bibitem{Horst13}
B. Horst, K. Bartkiewicz, and A. Miranowicz, \extra{Two-qubit
mixed states more entangled than pure states: Comparison of the
relative entropy of entanglement for a given nonlocality,} \pra
\textbf{87}, 042108 (2013).

\bibitem{Verstraete01}
F. Verstraete, K. M. R. Audenaert, J. Dehaene, and B. De Moor,
\extra{A comparison of the entanglement measures negativity and
concurrence,} J. Phys. A {\bf 34}, 10327 (2001).


\bibitem{Miran04}
A. Miranowicz and A. Grudka, \extra{Ordering two-qubit states with
concurrence and negativity,} \pra \textbf{70}, 032326 (2004);


\bibitem{Miran04b}
A. Miranowicz and A. Grudka, \extra{A comparative study of
relative entropy of entanglement, concurrence and negativity,} J.
Opt. B \textbf{6}, 542 (2004).


\bibitem{Miran08}
A. Miranowicz, S. Ishizaka, B. Horst, and A. Grudka,
\extra{Comparison of the relative entropy of entanglement and
negativity,} \pra \textbf{78}, 052308 (2008).


\bibitem{Augusiak08}
R. Augusiak, M. Demianowicz, and P. Horodecki, \extra{Universal
observable detecting all two-qubit entanglement and
determinant-based separability tests,} \pra {\bf 77}, 030301
(2008).


\bibitem{Bartkiewicz14drg}
K. Bartkiewicz, P. Horodecki, K. Lemr, A. Miranowicz, and K.
\.Zyczkowski, \extra{Method for universal detection of two-photon
polarization entanglement,}  e-print arXiv:1405.5560v1.

\bibitem{Bartkiewicz13discord}
K. Bartkiewicz, K. Lemr, A. \v{C}ernoch, and J. Soubusta,
\extra{Measuring nonclassical correlations of two-photon states,}
\pra {\bf 87}, 062102 (2013).

\bibitem{Bula13QND}
M. Bula, K. Bartkiewicz, A. \v{C}ernoch, and K. Lemr, 
\extra{Entanglement-assisted scheme for nondemolition detection
of the presence of a single photon,}
\pra{\textbf{87}}, 033826 (2013); E. Meyer-Scott, M. Bula, 
K. Bartkiewicz,  A. \v{C}ernoch, J. Soubusta, T. Jennewein, and K. Lemr
\extra{Entanglement-based linear-optical qubit amplifier,}
\pra {\bf 88}, 012327  (2013).


\bibitem{Bartkiewicz13fidelity}
K. Bartkiewicz, K. Lemr, and A. Miranowicz, \extra{Direct method
for measuring of purity, superfidelity, and subfidelity of
photonic two-qubit mixed states,} \pra {\bf 88}, 052104 (2013).


\bibitem{Peres96} 
A. Peres, \extra{Separability
Criterion for Density Matrices,} \prl \textbf{77}, 1413 (1996).

\bibitem{Bartkowiak11}
M. Bartkowiak, A. Miranowicz, X. Wang, Y.X. Liu, W. Leonski, and
F. Nori, \extra{Sudden vanishing and reappearance of nonclassical
effects: General occurrence of finite-time decays and periodic
vanishings of nonclassicality and entanglement witnesses,}
\pra~\textbf{83}, 053814 (2011).


\bibitem{Mintert07}
F. Mintert, \extra{Entanglement measures as physical observables,}
Appl. Phys. B \textbf{89}, 493 (2007).


\bibitem{Makhlin00}
Y. Makhlin, \extra{Nonlocal properties of two-qubit gates and
mixed states and optimization of quantum computations,} Quantum
Inf. Process. \textbf{1}(4),  243 (2002).


\bibitem{King06}
R. King and T. Welsh, \extra{Qubits and invariant theory,} J.
Phys.: Conf. Ser. \textbf{30}, 1 (2006).


\bibitem{Zyczkowski98}
K. \.Zyczkowski, P. Horodecki, A. Sanpera, and M. Lewenstein,
\extra{Volume of the set of separable states,} \pra \textbf{58},
883 (1998).

\bibitem{Vidal02}
G.~Vidal and R.~F. Werner, \extra{Computable measure of
entanglement,} \pra \textbf{65}, 032314 (2002).

\bibitem{Audenaert03}
K. Audenaert, M. B. Plenio, and J. Eisert, \extra{Entanglement
Cost under Positive-Partial-Transpose-Preserving Operations,} \prl
\textbf{90}, 027901 (2003).

\bibitem{Ishizaka04}
S. Ishizaka, \extra{Binegativity and geometry of entangled states
in two qubits,} \pra \textbf{69}, 020301(R) (2004).

\bibitem{Eltschka13}
C. Eltschka and J. Siewert, \extra{Negativity as an Estimator of
Entanglement Dimension,} \prl \textbf{111}, 100503 (2013).

\bibitem{Rana13}
S. Rana, \extra{Negative eigenvalues of partial transposition of
arbitrary bipartite states,} \pra \textbf{87}, 054301 (2013); A.
Sanpera, R. Tarrach, and G. Vidal, \extra{Local description of
quantum inseparability,} \textit{ibid.} \textbf{58}, 826 (1998).


\bibitem{Wootters98}
W. K. Wootters, \extra{Entanglement of Formation of an Arbitrary
State of Two Qubits,} \prl {\bf 80}, 2245 (1998).


\bibitem{Sinolecka12}
M. Sino\l{}\c{e}cka,  K. \.Zyczkowski, and M. Ku\'s,
\extra{Manifolds of equal entanglement for composite quantum
systems,} Acta Phys. Pol. B \textbf{33}, 2081 (2002); F.
Verstraete, J. Dehaene, and B. De Moor, \extra{Normal forms and
entanglement measures for multipartite quantum states,}  \pra
\textbf{68}, 012103 (2003); G. Gour, \extra{Family of concurrence
monotones and its applications,} {\it ibid.} \textbf{71}, 012318
(2005).


\bibitem{Werner89}
R.F. Werner, \extra{Quantum states with Einstein-Podolsky-Rosen
correlations admitting a hidden-variable model,} \pra \textbf{40},
4277 (1989).


\bibitem{Mendonca14}
P.E.M.F. Mendonca, M.A. Marchiolli, and D. Galetti,
\extra{Entanglement universality of two-qubit X-states,} e-print
arXiv:1407.3021.


\bibitem{Rigolin04}
G. Rigolin, \extra{Thermal entanglement in the two-qubit
Heisenberg XYZ model,} Int. J. Quant. Inf. \textbf{2}, 393 (2004).


\bibitem{Chen10}
T. Chen, Y.-X. Huang, C.-J. Shan, J.-X. Li, J.-B. Liu, and T.-K.
Liu, \extra{Entanglement evolution in an anisotropic two-qubit
Heisenberg XYZ model with Dzyaloshinskii-Moriya interaction,}
Chin. Phys. B \textbf{19}, 050302 (2010).


\bibitem{Yu04}
T. Yu and J. H. Eberly, \extra{Finite-Time Disentanglement Via
Spontaneous Emission,} Phys. Rev. Lett. \textbf{93}, 140404
(2004).


\bibitem{Ficek08}
Z. Ficek and R. Tana\'s, \extra{Dark periods and revivals of
entanglement in a two-qubit system,} Phys. Rev. A \emph{74},
024304 (2006); \extra{Delayed sudden birth of entanglement,}
\emph{ibid.} \textbf{77}, 054301 (2008).


\bibitem{Demianowicz11}
M. Demianowicz, \extra{Reexamination of determinant-based separability test for two qubits,} 
\pra \textbf{83}, 034301 (2011).

\bibitem{Yu07}
T. Yu and J.H. Eberly, \extra{Evolution from entanglement to
decoherence of bipartite mixed X states,} Quant. Inf. Comp.
\textbf{7}, 459 (2007).

\bibitem{Miran04a} 
A. Miranowicz, \extra{Violation of Bell inequality and 
entanglement of decaying Werner states,} \pla \textbf{327}, 
272 (2004). 

\bibitem{Hedemann13}
S.R. Hedemann, \extra{Evidence that All States Are Unitarily
Equivalent to X States of the Same Entanglement,}  e-print
arXiv:1310.7038v4.


\bibitem{Bennett96} C.~H. Bennett, G.~Brassard, S.~Popescu,
B.~Schumacher, J.~A. Smolin, and W.~K. Wootters, \extra{Purification of Noisy Entanglement and Faithful Teleportation via Noisy Channels,} Phys. Rev. Lett.
\textbf{76,} 722 (1996); C.~H. Bennett, D.~P. DiVincenzo, J.~A.
Smolin, and W.~K. Wootters, \extra{Mixed-state entanglement and quantum error correction,} Phys. Rev. A \textbf{54}, 3824 (1996).


\bibitem{Vedral98}
V. Vedral and M. B. Plenio, \extra{Entanglement measures and
purification procedures,} \pra {\bf 57}, 1619 (1998).



\end{thebibliography}
\end{document}